\documentclass[
    ,final            
    ,numberedheadings 
    ,cmfonts          
  ]
  {aipproc}
\usepackage{graphicx,epsfig,slashed}

\layoutstyle{6x9}

\begin{document}

\title{J/$\psi$ absorption in a multicomponent hadron gas}

\classification{14.40.Lb, 24.10.Pa, 25.75.-q, 12.38.Mh}
\keywords{Charmed mesons, Thermal and statistical models,
Relativistic Heavy-Ion Collisions, Quark Gluon Plasma}

\author{D.~Prorok}{
  address={Instytut Fizyki Teoretycznej,
Uniwersytet Wroc{\l}awski, 50-204 Wroc{\l}aw, Poland}
}

\author{L.~Turko}{
  address={Instytut Fizyki Teoretycznej,
Uniwersytet Wroc{\l}awski, 50-204 Wroc{\l}aw, Poland}
}

\author{D.~Blaschke}{
  address={Instytut Fizyki Teoretycznej,
Uniwersytet Wroc{\l}awski, 50-204 Wroc{\l}aw, Poland},
altaddress={Bogoliubov Laboratory for Theoretical Physics,
JINR, 141980 Dubna, Russia}
\\
E-mail: prorok@ift.uni.wroc.pl
}


\begin{abstract}
A model for anomalous $J/\Psi$ suppression in high energy heavy ion collisions
is presented.
As the additional suppression mechanism beyond standard nuclear absorption
inelastic $J/\Psi$ scattering with hadronic matter is considered.
Hadronic matter is modeled as an evolving multi-component gas of
point-like non-interacting particles (MCHG).
Estimates for the sound velocity of the MCHG are given and the equation of
state is compared with Lattice QCD data in the vicinity of the deconfinement
phase transition.
The approximate cooling pattern caused by longitudinal expansion is presented.
It is shown that under these conditions the resulting $J/\Psi$ suppression
pattern agrees well with NA38 and NA50 data.
\end{abstract}

\maketitle


\section {Introduction }
\label{intro}

The possible existence of a quark-gluon plasma (QGP) phase of highly excited
hadronic matter is one of the most intriguing questions of high energy physics
discussed since almost three decades up to now.
This novel state of matter has been also predicted in lattice QCD calculations
(for a review see \cite{Karsch:2004ik} and references therein) and the
critical temperature $T_{c}$ for the ordinary hadronic matter-QGP phase
transition has been obtained in the range of 150 - 270 MeV depending on the
number of active quark flavors
(this corresponds to the broad range of the critical energy densities
$\epsilon_{c} \simeq 0.26-5.5$ GeV/fm$^{3}$).
Since estimates of the NA50 collaboration for the energy density obtained in
the central rapidity region (CRR) give the value of 3.5 GeV/fm$^{3}$ for the
most central Pb-Pb data point, it has been argued that the
conditions for the existence of the QGP are met in Pb-Pb collisions at CERN
SPS \cite{Abreu:2000ni}.

The main argument for the QGP creation during Pb-Pb collisions at the CERN SPS
was the observation of the anomalous suppression of $J/\Psi$ relative yield.
$J/\Psi$ suppression as a signal for the QGP formation was originally proposed
by Matsui and Satz \cite{Matsui:1986dk}.
The idea is based on the modification of the quark forces in plasma due
to color screening which would entail a dissociation of  $c-\bar{c}$ bound
states, including the prominent $J/\Psi$ at high temperature, observable as
a significant depletion of the dilepton spectrum around the quarkonium
invariant mass.

The key point of the NA50 Collaboration paper \cite{Abreu:2000ni} was the
figure, denoted as Fig.6 there, where experimental data for Pb-Pb collision
values of ${B_{\mu\mu}\sigma_{J/\psi}} \over {\sigma_{DY}}$
(the ratio of the $J/\Psi$ to the Drell-Yan production cross-section times the
branching ratio of the $J/\Psi$ into a muon pair) were presented together with
some conventional predictions.
Here, "conventional" meant that $J/\Psi$ suppression was due to $J/\Psi$
absorption in ordinary hadronic matter. Since all those conventional curves
saturated at high transverse energy $E_{T}$, but the experimental data fell
from $E_{T} \simeq$ 90 GeV much lower and this behavior could be reproduced on
the base of $J/\Psi$ disintegration in the QGP \cite{Satz:1999si}, this was
argued to be circumstantial evidence for QGP formation in most central Pb-Pb
collisions.
In general, the immediate reservation about such reasoning was that besides
those already known (see e.g.
\cite{Ftacnik:1988qv,Gerschel:1988wn,Vogt:1988fj,Gavin:1988hs,Blaizot:1989ec,Vogt:1999cu} and
references [12-15] in \cite{Abreu:2000ni}), other possible models formulated
in terms of hadronic degrees of freedom could not be ruled out as long as the
$J/\Psi$ absorption rates in dense hadronic matter are largely unknown.

Within a few years the NA50 Collaboration has updated their $J/\Psi$ results
\cite{Ramello:2003ig}, using the nuclear absorption systematics obtained from
precise $pA$ results.
It appeared that also these data could be explained with a hadronic
comover absorption model \cite{Capella:2001vb}, without the necessity to
invoke QGP formation.
The question, however, arises whether the inputs used in these models are
consistent with accessible theoretical or experimental information.

In the present contribution, we present a systematic step towards and general
description of $J/\Psi$ absorption in the framework of a statistical analysis
which can provide a baseline in the search for non-hadronic explanations of
anomalous $J/\Psi$ suppression.

The main features of the model introduced in Ref. \cite{Prorok:2001kv} are

\begin{enumerate}
\item a multi-component non-interacting hadron gas appears in the CRR instead
of the QGP.
All hadrons from the lowest up to $\Omega^{-}$ baryon (with their non-zero
masses) are taken into account as constituents of the matter;
\item the gas expands longitudinally and transversely;
\item $J/\Psi$ suppression is the result of inelastic scattering on
constituents of the gas and on nucleons of colliding ions.
Both "traditional" sources of $J/\Psi$ suppression, namely absorption in the
nuclear matter and in the hadron gas in the CRR, are considered
simultaneously.
\end{enumerate}

The model has the following parameters: the initial time $t_{0}$, the
$J/\Psi$-baryon cross section $\sigma_{b}$, the initial baryon number density
$n_{B}^{0}$, $r_{0}$ in the expression $R_{A}=r_{0}A^{1 \over 3}$ and the
freeze-out temperature $T_{f.o.}$.
But the last quantity disappears effectively in the final estimations because
the "natural" freeze-out is enforced by the transverse expansion when the
rarefaction wave reaches the collision axis.

\section {The timetable of events in the CRR}
\label{timetable}

For a given A-B collision $t=0$ is fixed at the moment of the maximal overlap
of the nuclei (for more details see \emph{e.g.} \cite{Blaizot:1989ec}).
As the nuclei pass each other charmonium states are produced as the result of
gluon fusion.
After half of the time the nuclei need to cross each other ($t \sim$ 0.5 fm),
matter appears in the CRR.
It is assumed that the matter thermalizes almost immediately and the moment of
thermalization, $t_{0}$, is estimated to about 1 fm/c
\cite{Blaizot:1989ec,Bjorken:1983qr}.
Then the matter begins its expansion and cooling and after reaching the
freeze-out temperature, $T_{f.o.}$, it ceases as a thermodynamical system.
The moment when the temperature has decreased to $T_{f.o.}$ is denoted as
$t_{f.o.}$.
Since the matter under consideration is a gas of hadronic resonances, no
phase transition takes place during cooling.

For the description of the evolution of the matter, relativistic hydrodynamics
is employed.
The longitudinal component of the solution of the hydrodynamic equations (the
exact analytic solution for an (1+1)-dimensional case) reads (for details see,
e.g., \cite{Bjorken:1983qr,Cleymans:1986wb})
\begin{equation}
s(\tau)= { {s_{0}\tau_{0}} \over \tau } \;,\;\;\;\;\;\;\;\;\;
n_{B}(\tau)= { {n_{B}^{0}\tau_{0}} \over \tau }
\;,\;\;\;\;\;\;\;\;\; v_{z}={z \over t}
\label{bjork}
\end{equation}
%
where $\tau=\sqrt{t^{2}-z^{2}}$ is a local proper time, $v_{z}$ is the
component of the fluid velocity parallel to the collision axis and $s_{0}$ and
$n_{B}^{0}$ are the initial densities of the entropy and the baryon number,
respectively.
For $n_{B}=0$ and the uniform initial temperature distribution with a sharp
edge at the border established by the nuclear surfaces, the full solution of
the (3+1)-dimensional hydrodynamic equations is known \cite{Baym:1983sr}.
The evolution derived is decomposed into the longitudinal expansion inside
a slice bordered by the front of the rarefaction wave and the transverse
expansion which is superimposed outside the wave.
Since small but nevertheless non-zero baryon number densities are considered
here, the above-mentioned description of the evolution has to be treated as an
assumption in the presented model.
The rarefaction wave moves radially inward with the sound velocity $c_{s}$
(see Sect.~\ref{soundv}).

\section {The multi-component hadron gas }
\label{hadgas}

For an ideal multicomponent hadron gas in thermal and chemical equilibrium,
consisting of $l$ species (here, mesons are considered up to $K_{2}^{*}$ and
baryons up to $\Omega^{-}$), energy density $\epsilon$, baryon number density
$n_{B}$, strangeness density $n_{S}$ and entropy density $s$ are given by
($\hbar=c=1$ always)

\label{eqstate}
\begin{equation}
\epsilon = { 1 \over {2\pi^{2}}} \sum_{i=1}^{l} (2s_{i}+1)
\int_{0}^{\infty}dp\,{ { p^{2}E_{i} } \over { \exp \left\{ {{
E_{i} - \mu_{i} } \over T} \right\} + g_{i} } } \ , \label{energy}
\end{equation}

\begin{equation}
n_{B}={ 1 \over {2\pi^{2}}} \sum_{i=1}^{l} (2s_{i}+1)
\int_{0}^{\infty}dp\,{ { p^{2}B_{i} } \over { \exp \left\{ {{
E_{i} - \mu_{i} } \over T} \right\} + g_{i} } } \ ,
\label{barnumb}
\end{equation}

\begin{equation}
n_{S}={1 \over {2\pi^{2}}} \sum_{i=1}^{l} (2s_{i}+1)
\int_{0}^{\infty}dp\,{ { p^{2}S_{i} } \over { \exp \left\{ {{
E_{i} - \mu_{i} } \over T} \right\} + g_{i} } } \ ,
\label{strange}
\end{equation}

\begin{equation}
s={1 \over {6\pi^{2}T^{2}} } \sum_{i=1}^{l} (2s_{i}+1)
\int_{0}^{\infty}dp\, { {p^{4}} \over { E_{i} } } { { (E_{i} -
\mu_{i}) \exp \left\{ {{ E_{i} - \mu_{i} } \over T} \right\} }
\over { \left( \exp \left\{ {{ E_{i} - \mu_{i} } \over T} \right\}
+ g_{i} \right)^{2} } }\ , \label{entropy}
\end{equation}
%
where $E_{i}= ( m_{i}^{2} + p^{2} )^{1/2}$ and $m_{i}$, $B_{i}$, $S_{i}$,
$\mu_{i}$, $s_{i}$ and $g_{i}$ are the mass, baryon number, strangeness,
chemical potential, spin and a statistical factor of species $i$, respectively
(an antiparticle is treated as a different species).
The chemical potential $\mu_{i} = B_{i}\mu_{B} + S_{i}\mu_{S}$ defines that of
the overall baryon number, $\mu_{B}$ and that of strangeness, $\mu_{S}$.

To obtain the time dependence of temperature, baryon number and strangeness
chemical potentials one has to solve the equations (\ref{barnumb}) -
(\ref{entropy}) numerically with $s$, $n_{B}$ and $n_{S}$ given as time
dependent quantities.
For $s(\tau)$ and $n_{B}(\tau)$ the expressions (\ref{bjork}) are taken and
$n_{S}=0$ since the overall strangeness equals zero during the whole evolution,
see Sect.~\ref{power}.

\section {The sound velocity in the MCHG}
\label{soundv}

In the hadron gas the sound velocity squared is given by the standard
expression

\begin{equation}
c_{s}^{2}= { {\partial P} \over {\partial \epsilon} }\ . \label{speeds}
\end{equation}
%
Since the experimental data for heavy-ion collisions suggest that the baryon
number density is non-zero in the CRR at AGS and SPS energies
\cite{Baechler:1991pd,Stachel:1999rc,Ahle:1998jc}, we calculate the above
derivative for various values of $n_{B}$ \cite{Prorok:1995xz,Prorok:2001ut}.

\begin{figure}[hbt!]
\begin{tabular}{cc}
\includegraphics[width=0.45\textwidth,height=0.4\textwidth]{eps-hrg.eps}&
\includegraphics[width=0.45\textwidth,height=0.4\textwidth]{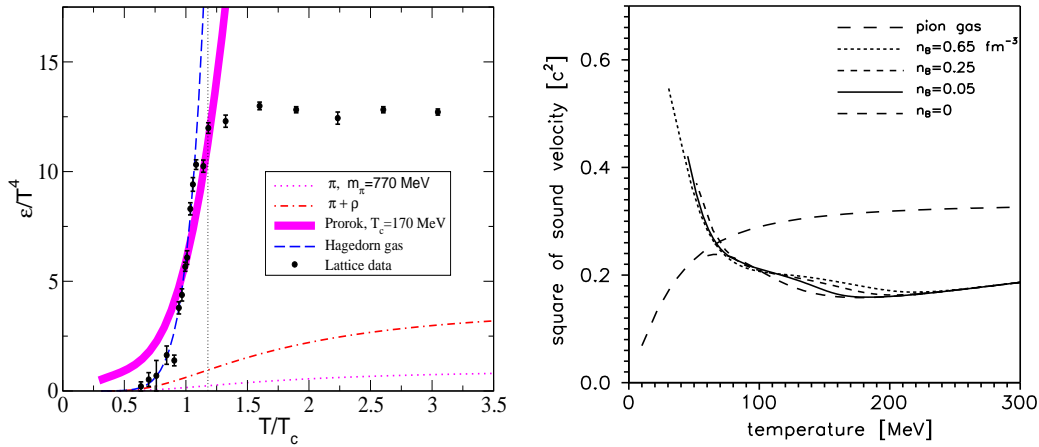}
\end{tabular}
\caption{
Left panel: Energy density in units of $T^4$ versus scaled temperature $T/T_c$
for the present hadron resonance gas model \protect\cite{Prorok:2001kv}
compared with the QCD Lattice data \protect\cite{Karsch:2000ps} and the
Hagedorn resonance gas with scaled hadron masses to adapt for the case of
unphysical quark masses in the Lattice simulation
\protect\cite{Karsch:2003vd}.
Right panel: Dependence of the sound velocity squared on temperature for
$n_{B}=0.65$ fm$^{-3}$ (short-dashed), $n_{B}=0.25$ fm$^{-3}$ (dashed),
$n_{B}=0.05$ fm$^{-3}$ (solid) and $n_{B}=0$ (long-dashed).
The case of the pure pion gas (long-long-dashed) is also presented.}
\label{Fig.1.}
\end{figure}

To estimate initial baryon number density $n_{B}^{0}$ we can use experimental
results for S-S \cite{Baechler:1991pd} or Au-Au
\cite{Stachel:1999rc,Ahle:1998jc} collisions.
In the first approximation we can assume that the baryon multiplicity per unit
rapidity in the CRR is proportional to the number of participating nucleons.
For S-S collisions we have $dN_{B}/dy \cong 6$ \cite{Baechler:1991pd} and 64
participating nucleons.
For central collisions of lead nuclei we can estimate the number of
participating nucleons as $2A = 416$, so we have $dN_{B}/dy \cong 39$.
Having taken the initial volume in the CRR equal to $\pi R_{A}^{2} \cdot 1$ fm,
we arrive at $n_{B}^{0} \cong 0.25$ fm$^{-3}$.
This is some underestimation because for S-S collisions the beam energy was at
200 GeV/nucleon, whereas for Pb-Pb at 158 GeV/nucleon.
From the Au-Au data extrapolation one can estimate $n_{B}^{0}\cong 0.65$
fm$^{-3}$ \cite{Stachel:1999rc}.
These values are for central collisions.
So, we estimate (\ref{speeds}) for $n_{B}= 0.25,\;0.65$ fm$^{-3}$ and
additionally, to investigate the dependence on $n_{B}$ much carefully, for
$n_{B}=0.05$ fm$^{-3}$.
The results of the numerical evaluation of (\ref{speeds}) are presented in
Fig.\,\ref{Fig.1.}.
For comparison, we show also curves for $n_{B}=0$ and for a pure massive pion
gas. These curves are taken from \cite{Prorok:1995xz}.

The "physical region" we consider between 100 and 200 MeV on the temperature
axis. This is because the critical temperature for the possible QGP-hadronic
matter transition is of the order of 200 MeV \cite{Karsch:2004ik} and the
freeze-out temperature should not be lower than 100 MeV \cite{Stachel:1999rc}.
At low temperatures we can see a completely different behaviour of the cases
with $n_{B}=0$ and $n_{B}\not=0$.
We think that this is caused by the fact that for $n_{B}\not=0$ the gas density
can not reach zero when $T\rightarrow 0$, whereas for $n_{B}=0$ it can.
For the higher temperatures all curves excluding the pion case behave
qualitatively in the same way.
From $T \approx 70$ MeV they decrease to their minima (for $n_{B}=0.65$
fm$^{-3}$ at $T \cong 219.6$ MeV, for $n_{B}=0.25$ fm$^{-3}$ at
$T \cong 202.5$ MeV, for $n_{B}=0.05$ fm$^{-3}$ at $T \cong 183.4$ MeV and for
$n_{B}=0$ at $T \cong 177.3$ MeV) and then they increase to merge to each other
above $T \approx 250$ MeV.

\section {The cooling pattern vs. sound velocity}
\label{power}

In Sect.~\ref{hadgas} we have explained how to obtain the time dependence of
the temperature of the longitudinally expanding hadron gas.
This dependence proved to be very well approximated by the expression
\cite{Prorok:1995xz,Prorok:2001ut}

\begin{equation}
T(\tau) \cong T_{0} \cdot \tau^{-a}\ .
\label{cooling}
\end{equation}
%
The above approximation is valid in the temperature range $[T_{f.o.},T_{0}]$,
where $T_{f.o.} \geq 100$ MeV, $T_{0} \leq T_{0,max}$ and
$T_{0,max} \approx 230$ MeV.
We started from $T_{0}$ equal to 227.2 MeV (for $n_{B}^{0} = 0.65$ fm$^{-3}$),
229.3 MeV (for $n_{B}^{0} = 0.25$ fm$^{-3}$) and 229.7 MeV
(for $n_{B}^{0} = 0.05$ fm$^{-3}$).
These values correspond to $\epsilon_{0} = 5.0$ GeV/fm$^{3}$ (the initial
energy density in the CRR has been estimated by NA50 \cite{Abreu:2000ni} to
$\epsilon_{0} = 3.5$ GeV/fm$^{3}$).
Then we took several decreasing values of $T_{0} < T_{0,max}$.
For every $T_{0}$ chosen we repeat the procedure of obtaining the approximation
(\ref{cooling}), i.e. the coefficient $a$ of the power law.


We can formulate the following conclusion: in the "physical region" of
temperature and for realistic baryon number densities, the longitudinal
expansion given by (\ref{bjork}) results in the cooling of the hadron gas
described by (\ref{cooling}) with $a=c_{s}^2(T_{0})$, namely:

\begin{equation}
T(\tau) \cong T_{0} \cdot \left( {\tau_{0} \over \tau}
\right)^{c_{s}^{2}(T_{0})}\,
\label{coolapp}
\end{equation}
%
where $T_{0}$ belongs to the "physical region". Note that
$T(\tau) = T_{0} \cdot \left( {\tau_{0} \over \tau} \right)^{c_{s}^{2}}$ is the
exact expression for a baryonfree gas with the sound velocity constant
(for details see \cite{Cleymans:1986wb,Baym:1983sr}).
It should be stressed that $c_{s}^{2}$ in (\ref{coolapp}) is constant depending
only on the initial temperature $T_{0}$ from which the cooling starts.
The approximation (\ref{coolapp}) will be used to simplify the evaluation of
the $J/\Psi$ survival factors in the next sections.

It should be added that for the excluded volume hadron gas model the
above-mentioned conclusion is no longer valid.
We perfprmed the appropriate simulations, but approximations of $T(\tau)$ by a
power law turned out to be rather inaccurate and yielded a coefficient $a$
which was by a factor two smaller than the squared velocity of sound.

\section {$J/\Psi$ absorption in the expanding MCHG }
\label{absorb}

As it has been already mentioned in Sect.~\ref{timetable}, charmonium states
are produced in the beginning of the collision, when nuclei overlap.
For simplicity, it is assumed that the production of $c\bar{c}$ states takes
place at $t=0$.
To describe $J/\Psi$ absorption quantitatively, the idea of Ref.
\cite{Blaizot:1989ec} is generalized here to the case of the multi-component
massive gas.
Since in the CRR longitudinal momenta of particles are much lower than
transverse ones (in the c.m.s. frame of nuclei), the $J/\Psi$ longitudinal
momentum is put to zero.
Additionally, only the plane $z=0$ is under consideration.
For the simplicity of the model, it is assumed that all charmonium states are
completely formed and can be absorbed by constituents of the surrounding medium
from the moment of their creation by inelastic scattering through interactions
of the type
\begin{equation}
c\bar c+h \longrightarrow D+\bar D+X\\ ,  \label{psiabs}
\end{equation}
%
where $h$ denotes a hadron, $D$ is a charm meson and $X$ stands for a particle
which is necessary to conserve the charge, baryon number or strangeness.

Since Pb-Pb collisions are the most relevant case for the problem of QGP
search (see remarks in Sect.~\ref{intro}), the further considerations are done
for this case.

%
It is assumed that the hadron gas, which appears in the space between the
nuclei after they have crossed each other, also has the shape of the overlap
area of the colliding nuclei ($S_{eff}$) at $t_{0}$ in the $z=0$ plane.
Then, the transverse expansion starts as a rarefaction wave moving inward
$S_{eff}$ at $t_{0}$.
From the considerations based on the relativistic kinetic equation (for details
see \cite{Prorok:2001kv,Blaizot:1989ec}), the survival fraction of $J/\Psi$ in
the hadron gas as a function of the initial energy density $\epsilon_{0}$ in
the CRR is obtained:
\begin{equation}
{\cal N}_{h.g.}(\epsilon_{0}) = \int dp_{T}\;
g(p_{T},\epsilon_{0}) \cdot \exp \left\{ -\int_{t_{0}}^{t_{final}}
dt \sum_{i=1}^{l} \int { {d^{3}\vec{q}} \over {(2\pi)^{3}} }
f_{i}(\vec{q},t) \sigma_{i} v_{rel,i} { {p_{\nu}q_{i}^{\nu}} \over
{EE^{\prime}_{i}} } \right\}\ ,
\label{surhg}
\end{equation}
%
where the sum is over all species of scatters (hadrons),
$p^{\nu}=(E,\vec{p}_{T})$ and $q_{i}^{\nu}=(E^{\prime}_{i},\vec{q})$ are four
momenta of $J/\Psi$ and hadron species $i$ respectively, $\sigma_{i}$ stands
for the absorption cross-section of $J/\Psi-h_{i}$ scattering, $v_{rel,i}$ is
the relative velocity of  $J/\Psi$ and hadron $h_{i}$. $M$ and
$m_{i}$ denote $J/\Psi$ and $h_{i}$ masses respectively ($M= 3097$ MeV).
The function $g(p_{T},\epsilon_{0})$ is the $J/\Psi$ initial momentum
distribution.
It has a gaussian form and reflects gluon multiple elastic scattering on
nucleons before their fusion into a $J/\Psi$ in the first stage of the
collision \cite{Hufner:1988wz,Gavin:1988tw,Blaizot:1989hh}.
The upper limit $t_{final}$  of the time integration in (\ref{surhg}) is the
minimal value of $\langle t_{esc}\rangle$ and $t_{f.o.}$.
The quantity $\langle t_{esc}\rangle$ is the average time of the escape of
$J/\Psi$'s from the hadronic medium for given values of $b$ and $J/\Psi$
velocity $\vec{v}= \vec{p}_{T}/E$.
Note that the average is taken with the weight
\begin{equation}
p_{J/\Psi}(\vec{r}) = T_{A}(\vec{r})T_{B}(\vec{r} - \vec{b})/T_{AB}(b)~,
\label{weight}
\end{equation}
%
where
$T_{AB}(b) = \int d^{2}\vec{s}\; T_{A}(\vec{s}) T_{B}(\vec{s} - \vec{b})$,
$T_{A}(\vec{s}) = \int dz \rho_{A}(\vec{s},z)$
and
$\rho_{A}(\vec{s},z)$
is the nuclear matter density distribution taken here as the Woods-Saxon form
with parameters from \cite{Jager}.
In the integration over hadron momentum in (\ref{surhg}) the threshold for the
reaction (\ref{psiabs}) is included, i.e. $\sigma_{i}$ equals zero for
$(p^{\nu}+q_{i}^{\nu})^{2} < (2m_{D} + m_{X})^{2}$ and is constant elsewhere
($m_{D}$ is a charm meson mass, $m_{D}= 1867$ MeV).
Also the usual Bose-Einstein or Fermi-Dirac distribution for hadron species
$i$ is used in (\ref{surhg})
\begin{equation}
f_{i}(\vec{q},t)=f_{i}(q,t)=(2s_{i}+1) \bigg\{ \exp
\left[{E^{\prime}_{i}-\mu_{i}(t)} \over {T(t)} \right] + g_{i} \bigg\}^{-1}\ .
\label{BoseF}
\end{equation}
%
For simplicity, we use (\ref{coolapp}) as the approximation to ${T(t)}$ in
(\ref{BoseF}) and $\mu_{B}(t)$ and $\mu_{S}(t)$ are solutions of only two
equations (\ref{barnumb}) and (\ref{strange}) with ${T}$ given by
(\ref{coolapp}), $n_{B}(t)$ by (\ref{bjork}) and $n_{S}(t)=0$.

As far as $\sigma_{i}$ is concerned, there are no data for every particular
$J/\Psi-h_{i}$ scattering.
Therefore, we use here universal, energy independent cross sections for
 scattering of charmonia on baryons, $\sigma_{b}$, and on mesons,
$\sigma_m=2 \sigma_{b}/3$, according to the quark counting rules.

As it has been already suggested \cite{Gerschel:1988wn} also $J/\Psi$
scattering in nuclear matter should be included in any $J/\Psi$ absorption
model.
This could be done with the introduction of a $J/\Psi$ survival factor in
nuclear matter \cite{Gerschel:1992uh}
%
\begin{equation}
{\cal N}_{n.m.}(\epsilon_{0}) \cong \exp \left\{ -\sigma_{J/\psi N}
\rho_{0} L \right\}\ , \label{surnm}
\end{equation}
%
where $\rho_{0}$ is the nuclear matter density and $L$ the mean path length of
the $J/\Psi$ through the colliding nuclei obtained according to
%
\begin{equation}
\rho_{0}L(b) = {1 \over {2 T_{AB}} } \int d^{2}\vec{s}\;
T_{A}(\vec{s}) T_{B}(\vec{s} - \vec{b}) \left[ {{A-1} \over A}
T_{A}(\vec{s}) + {{B-1} \over B} T_{B}(\vec{s} - \vec{b})
\right]\ . \label{length}
\end{equation}
%
Since the $J/\Psi$ absorption processes in nuclear matter and in the MCHG are
separated in time, the $J/\Psi$ survival factor for a heavy-ion collision with
the initial energy density $\epsilon_{0}$, could be defined as
\begin{equation}
{\cal N}(\epsilon_{0}) = {\cal N}_{n.m.}(\epsilon_{0}) \cdot {\cal
N}_{h.g.}(\epsilon_{0})\ . \label{surv}
\end{equation}
%
Note that since the right hand sides of Eqs. (\ref{surhg}) and (\ref{surnm})
include parts which depend on the impact parameter $b$ and the left hand sides
are functions of $\epsilon_{0}$ only, the expression converting the first
quantity to the second (or reverse) should be defined.
This is done with the use of the dependence of $\epsilon_{0}$ on the
transverse energy $E_{T}$ extracted from NA50 data \cite{Abreu:2000ni}
(for details see \cite{Prorok:2001kv}).

To make the model as realistic as possible, one should keep in mind that
only about $60 \%$ of the observed $J/\Psi$'s are directly produced during the
collision.
The remainder is the result of $\chi$ ($\sim 30 \%$) and $\psi'$ ($\sim 10 \%$)
decays.
Therefore the realistic $J/\Psi$ survival factor could be expressed as
\begin{equation}
{\cal N}(\epsilon_{0})=0.6{\cal
N}_{J/\psi}(\epsilon_{0})+0.3{\cal N}_{\chi}(\epsilon_{0})+
0.1{\cal N}_{\psi'}(\epsilon_{0})\;, \label{survsum}
\end{equation}
%
where ${\cal N}_{J/\psi}(\epsilon_{0})$, ${\cal N}_{\chi}(\epsilon_{0})$ and
${\cal N}_{\psi'}(\epsilon_{0})$ are given also by Eqs. (\ref{surhg}),
(\ref{surnm}) and (\ref{surv}) but with $g(p_{T},\epsilon_{0})$,
$\sigma_{J/\psi N}$ and $M$ changed appropriately (for details see
\cite{Prorok:2001kv}).


\section { Results }
\label{result}

\begin{figure}[hbt!]
\includegraphics[width=0.45\textwidth,height=0.4\textwidth]{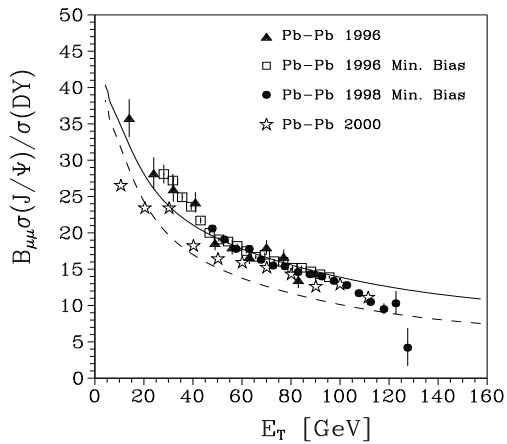}
\caption{
$J/\Psi$ survival factor times
${B_{\mu\mu}\sigma_{J/\psi}^{pp}}/{\sigma_{DY}^{pp}}$ in the
longitudinally and transversely expanding hadron gas for the Woods-Saxon
nuclear matter density distribution and $n_{B}^{0}=0.25$ fm$^{-3}$,
$T_{f.o.}=140$ MeV, $c_{s}=0.45$ and $r_{0}=1.2$ fm.
The curves correspond to $\sigma_{b}=4$ mb (solid) and $\sigma_{b}=5$ mb
(dashed).
The black triangles represent the 1996 NA50 Pb-Pb data, the white squares the
1996 analysis with minimum bias, the black points the 1998 analysis with
minimum bias \protect\cite{Abreu:2000ni}, and the white stars to the 2000
NA50 Pb-Pb reanalyzed data \protect\cite{Ramello:2003ig}. }
\label{Fig.9.}
\end{figure}

Now we can complete the calculations of formula (\ref{surhg}) for values of
$n_{B}^{0}$ given in Sect.~\ref{soundv}. The last parameter of the model is
the freeze-out temperature.
Two values $T_{f.o.}=100, 140$ MeV are taken here and they agree well with
estimates based on hadron yields \cite{Stachel:1999rc}.
The main results are presented in Fig.\,\ref{Fig.9.}.
The original data \cite{Abreu:2000ni} for
${B_{\mu\mu}\sigma_{J/\psi}^{PbPb}}/{\sigma_{DY}^{PbPb}}$ and
$J/\Psi$ survival factors given by (\ref{survsum}) multiplied by
${B_{\mu\mu}\sigma_{J/\psi}^{pp}}/ {\sigma_{DY}^{pp}}$  as functions of
$E_{T}$ are depicted there (for details see \cite{Prorok:2001kv}).
%
When comparing the present model with the data from the NA50 collaboration
in Fig.~\ref{Fig.9.}, we want to focus on the region below $E_T\sim 100$ GeV,
which corresponds to initial energy densities below $\varepsilon_0\sim 3.0$
GeV/fm$^3$ ($T_0<210$ MeV), where the MCHG gives an acceptable description
of the equation of state for hot and dense matter, in agreement with lattice
QCD data, see the left panel of Fig.~\ref{Fig.1.}.
In our phenomenological analysis shown in Fig.~\ref{Fig.9.}, we assume
universal cross sections for mesons and baryons, with the appropriate
thresholds for their dissociation reactions but energy-independent,
$\sigma_b=4$ mb (solid line) and $5$ mb (dashed line).
Note that the solid line perfectly describes the data between the onset
of the deviation from the nuclear absorption baseline at  $E_T\sim 40$ GeV
and the limit of applicability of the MCHG picture at  $E_T\sim 100$ GeV.
Does this result disprove the claim of the NA50 collaboration that the
onset of anomalous suppression at $E_T\sim 40$ GeV is evidence for the creation
of a new form of matter, made of deconfined quarks and gluons (QGP) in this
experiment?

Such a claim would be premature. The weak point is the question about the
charmonium dissociation cross sections. Although the absorption cross section
for $J/\psi$ in cold nuclear matter has been measured in $pA$ collisions
by the NA50 experiment to be $\sigma_{J/\psi N}=4.2 \pm 0.5$ mb
\cite{Alessandro:2003pc}, this
may be a result of the downfeeding from $\chi_c$ and $\psi'$ with larger cross
sections, so that the ``true'' absorption cross section for the ground state
component is between 2 and 3 mb, see also \cite{He:1999aj}.
No experimental information for the charmonia absorption cross sections on
higher baryonic resonances or on mesons is available yet.

However, there is progress in theoretical approaches.
Based on a diagrammatic approach to quark exchange processes in
hadron-hadron scattering \cite{Martins:1994hd}, the
dissociation cross section for charmonia on mesons have been calculated
\cite{Barnes:2003dg} to have a sharp rise at threshold to maxima between
1 ad 10 mb, depending on the channel, followed by a fast decrease in energy
due to vanishing overlap integrals between asymptotic mesonic states.
A recent calculation within a fully relativistic approach has confirmed the
result for $J/\psi$ dissociation by pion impact \cite{Ivanov:2003ge}.
Similar calculations for baryon impact  \cite{Hilbert:2007hc} show that both
assumptions made in this work could be disproved: the universality and energy
independence.

\section { Conclusions }
\label{conclu}

Properties of the hadron resonance gas model under conditions reached in the
ultrarelativistic heavy ion experiments are very close to the phase transition
region estimated from the lattice simulation data, see Fig.\,\ref{Fig.1.}.
%
We have shown that with $J/\Psi$ absorption cross sections on hadrons of
$4$ mb, an overall satisfactory description of NA38/NA50 data on
$J/\Psi$ suppression could be given here.
Therefore, one could argue that the NA50 Pb-Pb data do not provide evidence
for the production of {\it deconfined} matter in the central
rapidity region of a Pb-Pb collision.

 It is an interesting task for forthcoming work to use those energy dependent,
non-universal cross sections of the quark exchange model in the calculation
of charmonium dissociation in the MCHG model.
We expect that the outcome of such a calculation will
leave room for the discussion of in-medium modification (increase) of
dissociation rates by changes in the spectrum (broadening) of the final state
open charm hadrons which determine (effectively lower) the reaction thresholds.
Examples have been given for the $\pi-\rho$ gas \cite{Blaschke:2002ww} and
for nuclear matter \cite{Tsushima:2000cp}.
A heuristic extension of the MCHG to include spectral broadening due to the
Mott dissociation of hadrons (Mott-Hagedorn resonance gas) has been given in
\cite{Blaschke:2004xg}, but deserves a microscopic foundation.
%
Therefore, we conclude that the hadronic absorption cross sections of
the $J/\Psi$
need to be determined
to a higher accuracy before the anomalous $J/\Psi$ suppression could be
interpreted as a good signal for QGP
formation in central heavy-ion collisions.
\subsection*{Acknowledgments}
This work was supported in part by the Polish Ministry of Science and  Higher
Education under contract No.~N~N202~0953~33.

\end{document}